\documentclass[aps,prl,superscriptaddress,showpacs,twocolumn]{revtex4}

\usepackage{graphicx}
\usepackage{psfrag}
\usepackage{dcolumn}
\usepackage{bm}

\begin{document}

\title{Stability and phase coherence of trapped 1D Bose gases}

\author{D.M.~Gangardt}
\affiliation{Laboratoire Kastler-Brossel, Ecole Normale Sup\'erieure, 24 rue
Lhomond, 75005 Paris, France}
\author{G.V.~Shlyapnikov}
\affiliation{Laboratoire Kastler-Brossel, Ecole Normale Sup\'erieure, 24 rue
Lhomond, 75005 Paris, France}

\affiliation{\mbox{FOM Institute for Atomic and Molecular Physics, Kruislaan
407, 1098 SJ Amsterdam, The Netherlands}}

\affiliation{Russian Research Center Kurchatov Institute, Kurchatov
  Square, 123182 Moscow, Russia}

\date{\today}

\begin{abstract}
We discuss stability and phase coherence of 1D trapped Bose gases 
and find that inelastic decay processes, such as three-body recombination,
are suppressed in the strongly interacting (Tonks-Girardeau) and intermediate
regimes. This is promising for achieving these regimes with a large
number of particles. ''Fermionization'' of the system reduces the
phase coherence length, and at $T=0$ the gas is fully phase coherent
only deeply in the weakly interacting (Gross-Pitaevskii) regime. 

\end{abstract}

\pacs{03.75Fi, 05.30Jp}

\maketitle

Recent success in creating quantum degenerate 1D trapped atomic gases
\cite{Goerlitz2001,Schreck2001,Greiner2001} has stimulated an interest in
correlation properties of these systems. The 1D regime is reached by tightly
confining the radial motion of atoms in a cylindrical trap to zero point
oscillations.  The difference between such a kinematically 1D gas and a purely
1D gas is related only to the value of the interparticle interaction, which
depends on the radial confinement.

The 1D Bose gas with repulsive short-range interactions characterized by the
coupling constant $g>0$ exhibits remarkable properties. Counterintuitively, it
becomes more non-ideal with decreasing 1D density $n$
\cite{Girardeau1960,LiebLiniger1963}. The equation of state and correlation
functions depend crucially on the parameter
\begin{equation}
\label{eq:gamma}
\gamma=mg/\hbar^2 n,
\end{equation}
where $m$ is the atom mass. For comparatively large $n$, the parameter
$\gamma\ll 1$ and $\sqrt{\gamma}$ represents the ratio of the mean
interparticle separation $1/n$ to the correlation length
$l_c=\hbar/\sqrt{mng}$. In this case the gas is in the weakly interacting or
Gross-Pitaevskii (GP) regime, and the amplitude of the boson field obeys the
familiar Gross-Pitaevskii equation. For sufficiently low densities (or large
$g$), one has $\gamma\gg 1$ and the strongly interacting or Tonks-Girardeau
(TG) regime is reached. In this regime the 1D gas acquires fermionic
properties in the sense that the wave function strongly decreases as particles
approach each other (see \cite{Girardeau1960,LiebLiniger1963} and cf.
\cite{Tonks1936}). For $\gamma=\infty$ any correlation function of the density
can be calculated straightforwardly \cite{MehtaRandMatr} by using the exact
mapping onto the system of free fermions found by Girardeau
\cite{Girardeau1960}.

A uniform 1D system of bosons interacting via a delta-functional potential is
integrable by using the Bethe ansatz and has been a subject of extensive
theoretical studies. Lieb and Liniger \cite{LiebLiniger1963} have calculated
the ground state energy and excitation spectrum for any value of the parameter
$\gamma$. Yang and Yang \cite{Yang1969} have studied thermodynamic properties
of this system at finite temperatures and found no phase transition for $T>0$.
The absence of long-range order and true Bose-Einstein condensate in
finite-temperature 1D Bose gases follows from the Bogoliubov $k^{-2}$ theorem
as has been expounded in \cite{Mermin1966}. The long-range order is destroyed
by long-wave fluctuations of the phase leading to an exponential decay of the
one-particle density matrix at large distances \cite{Kane1967,Popov}.  A
similar picture is found at $T=0$ \cite{Pitaevskii1991}, where the density
matrix undergoes a power-law decay \cite{Schwartz1977,Haldane1981,Popov}. In
the past few decades, a general approach has been developed for exact
calculations of one- and two-particle correlation functions at an arbitrary
$\gamma$ (see \cite{KorepinBook,Thacker1981} for review).

Realization of 1D trapped Bose gases raises the question of their phase
coherence and stability. The phase coherence properties are strongly
influenced by the trapping potential, which introduces a finite size of the
system and provides a low-momentum cut-off of the phase fluctuations.  In the
GP regime at sufficiently low $T$, the phase fluctuations are suppressed and
the equilibrium state is a true condensate. At higher temperatures, it is
transformed into a phase-fluctuating condensate \cite{Petrov2000}. The
dynamics, interference effects, and excitations of 1D trapped Bose gases are
currently a subject of active studies
\cite{Girardeau2000,Dunjko2001,Stringari2002,Ohberg2002}. Of particular
interest is the change in phase coherence properties while going from the GP
to the TG regime, where the phase coherence is completely lost.

The strong transverse confinement required for the 1D regime can lead to high
3D densities of a trapped gas. At a large number of particles, the 3D density
can exceed $10^{15}$~cm$^{-3}$, and one expects a fast decay due to three-body
recombination. It is then crucial to understand how the correlation properties
of the gas influence the decay rate.

In this Letter, we discuss stability of 1D Bose gases and calculate local
density correlators, as those are responsible for inelastic decay processes
\cite{Kagan1985}. We find that the decay rates are suppressed in the TG and
intermediate regimes, which is promising for achieving these regimes with a
large number of particles. We then analyze phase coherence of a trapped 1D
Bose gas and show that vacuum fluctuations of the phase make the
zero-temperature coherence length smaller than the Thomas-Fermi size of the
sample, unless the gas is deeply in the GP regime.

The 1D regime in a trapped gas is realized if the amplitude of transverse zero
point oscillations $l_0=\sqrt{\hbar/m\omega_0}$ is much smaller than the
longitudinal correlation length $l_c=\hbar/\sqrt{m\mu}$, where $\omega_0$ is
the frequency of the transverse confinement and the chemical potential of the
1D system is $\mu\ll \hbar\omega_0$. One then has a 1D system of bosons
interacting with each other via a short-range potential characterized by an
effective coupling constant $g>0$.  This constant is expressed through the 3D
scattering length $a$ \cite{Olshanii98}, assuming that $l_0$ greatly exceeds
the radius of interaction between atoms. For a positive $a\ll l_0$ we have
\begin{equation}
\label{eq:geff}
g=2\hbar^2 a/m l_0^2,
\end{equation}
and a characteristic distance $\hbar^2/mg$ related to the interaction between
particles in the described 1D problem is $\sim l_0^2/a\gg l_0$.  In the GP
regime, the chemical potential $\mu\approx gn$, and the condition $l_c\gg l_0$
leads to the inequality $na\ll 1$. In the TG regime the correlation length
$l_c\sim 1/n$, and one should have $nl_0\ll 1$.  We thus see that at any value
of $\gamma$ it is sufficient to satisfy the inequalities $a\ll l_0\ll 1/n$.
Then the 1D regime is reached and the system can be analyzed on the basis of
the 1D Lieb-Liniger model, assuming a delta-functional interatomic potential
with the coupling constant $g$ given by Eq.~(\ref{eq:geff}).

The rate of three-body recombination is proportional to the local
three-particle correlation function $g_3=\langle\Psi^\dagger(x)\Psi^\dagger(x)
\Psi^\dagger(x)\Psi(x)\Psi(x)\Psi(x)\rangle$ \cite{Kagan1985}, where $\Psi
(x)$ is the field operator of atoms and the symbol $\langle\ldots\rangle$
denotes the expectation value. Similarly, the rates of two-body inelastic
processes involve the correlation function $g_2
=\langle\Psi^\dagger(x)\Psi^\dagger(x)\Psi(x)\Psi(x)\rangle$.  Assuming that
local correlation properties are insensitive to the geometry of the system we
consider a uniform 1D gas of $N$ bosons on a ring of circumference $L$.  The
Hamiltonian of the system reads:
\begin{equation}
\label{eq:ham}
H=\int dx\left[(\hbar^2/2m)\partial_x\Psi^\dagger\partial_x\Psi+
(g/2)\Psi^\dagger\Psi^\dagger\Psi\Psi\right].
\end{equation}
 
\begin{figure}
  \begin{center}
    \psfrag{xlabel}{\large$\gamma$} \psfrag{ylabel}{\large$g_2(\gamma)/n^2$}
    \psfrag{0}{\small 0} \psfrag{1}{\small 1}   
    \psfrag{5}{\small 5}    \psfrag{10}{\small 10}
    \psfrag{15}{\small 15}    \psfrag{20}{\small 20} 
    \includegraphics[
width=3.375in,
height=1.6in]{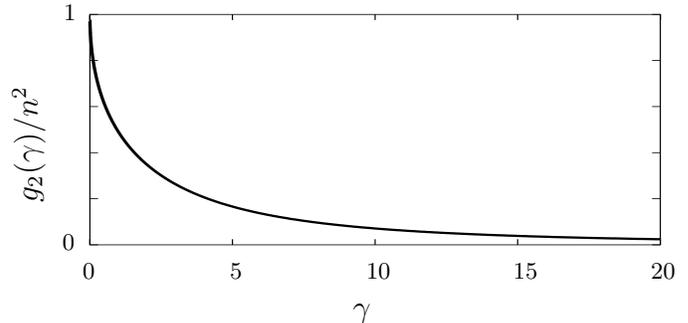}
\vspace{-0.7cm}
    \caption{Local correlation function $g_2$ versus $\gamma$.}
\vspace{-0.6cm}
    \label{fig:phaseterm.1}
  \end{center}
\end{figure}

For finding $g_2$ at $T=0$, we use the Hellmann-Feynman theorem
\cite{Hellmann1933,Feynman1939}. This is similar to the calculation of the
mean interaction energy in ref. \cite{LiebLiniger1963}.  Namely, one shows
that the expectation value of the four-operator term in the Hamiltonian
(\ref{eq:ham}) is proportional to the derivative of the ground state energy
with respect to the coupling constant: $ d E_0/d g = \langle \Phi_0 | d H/dg
|\Phi_0\rangle = g_2L/2$. The first identity follows from the normalization of
the ground state wave function $\Phi_0$, and the second one is obtained
straightforwardly from the Hamiltonian (\ref{eq:ham}). The ground state energy
can be written as $E_0 = N e(\gamma) \hbar^2 n^2 /2m$, where the quantity
$e(\gamma)$ is a solution of the Lieb-Liniger equations \cite{LiebLiniger1963}
and is calculated numerically for any value of $\gamma$
\cite{Dunjko2001,Chiara2002}. For the two-particle local correlation function
we then obtain $g_2(\gamma) =n^2de(\gamma)/d\gamma$.  The function $g_2
(\gamma)/n^2$ is shown in Fig.~\ref{fig:phaseterm.1}.  For small values of
$\gamma$, we obtain numerically the result which coincides with that following
from the Bogoliubov approach:
\begin{equation}
\label{eq:g2small}
   g_2 (\gamma)/n^2= 1-2\sqrt{\gamma}/\pi,\;\;\;\;\gamma\ll 1.
\end{equation}
In the limit of large $\gamma$, we have $e(\gamma)=
\left(\pi^2/3\right)\left(1-4/\gamma\right)$, and the two-particle correlation
function is given by
\begin{equation}
  \label{eq:g2large}
   g_2 (\gamma)/n^2= 4\pi^2/3\gamma^2,\;\;\;\;\;
   \gamma\gg 1.
\end{equation}
The results in Fig.~\ref{fig:phaseterm.1} and Eq.~(\ref{eq:g2large}) clearly
show that two-particle correlations and, hence, the rates of pair inelastic
processes, are suppressed for $\gamma \agt 1$. This provides a possibility
for identifying the TG and intermediate regimes of a trapped 1D Bose gas
through the measurement of photoassociation in pair interatomic
collisions.

The three-particle local density correlator $g_3$ cannot be obtained from the
Hellmann-Feynman theorem. In the weak coupling GP regime ($\gamma\ll 1$) one
can use the Bogoliubov approach, which immediately yields
\begin{equation}
\label{g3small}
 g_3 (\gamma)/n^3\simeq 1-6\sqrt{\gamma}/\pi,\;\;\;\;\gamma\ll 1.
\end{equation}

For the TG regime ($\gamma\gg 1$), we have developed a method
for calculating the leading behavior of local density correlators. Details 
will be given elsewhere, and here we present a compact derivation
of $g_3$ at $T=0$. In first quantization
the expression for this function reads
\begin{eqnarray}
\label{eq:g3first}
g_3(\gamma)=\frac{N!}{3!(N-3)!}\int
dX\left|\Phi^{(\gamma)}_0(0,0,0,x_4,\ldots,x_N)\right|^2, 
\end{eqnarray}
where $dX=dx_4\ldots dx_N$, and $\Phi^{(\gamma)}_0$ is the ground state 
wave function given in the domain
$0<x_1<\ldots<x_N<L$ by the Bethe ansatz solution:
\begin{eqnarray}
\Phi^{(\gamma)}_0(x_1,x_2,\ldots,x_N)\propto
\sum_{P}a(P)\exp\left\{i\sum\!k_{P_j}x_j\right\},
\label{eq:gsfunc}
\end{eqnarray}
where $P$ is a permutation of $N$ numbers, quasimomenta $k_j$ are solutions
of the Bethe ansatz equations, and
\begin{eqnarray}
a(P) = \prod_{j<l} \left(\frac{i\gamma n + k_{P_j}-k_{P_l}}
{i\gamma n - k_{P_j}+k_{P_l}}\right)^\frac{1}{2}.  \nonumber
\end{eqnarray}
For $\gamma\gg 1$, we extract the leading contribution to $\Phi_0^{(\gamma)}$
at three coinciding points by symmetrizing the amplitudes $a(P)$ over the first
three elements of the permutation $P$:
\begin{eqnarray}
\label{eq:apsym}
\frac{1}{3!}\sum_p a(P_p)\simeq
\frac{\epsilon_P}{(i\gamma n)^3}
\prod_{j<l}(k_{P_j}-k_{P_l}),    
\end{eqnarray} 
where $P_p=P_{p_1},P_{p_2},P_{p_3},P_4,\ldots P_N$, and $j,l=1,2,3$. 
The sign of the permutation $P$ is $\epsilon_P$, and $p$
runs over six permutations of $1,2,3$. For large $\gamma$, the difference of
quasi-momenta $k_j$ from their values at
$\gamma=\infty$  is of order $1/\gamma$ and can be neglected. 
Then, from Eqs.~(\ref{eq:gsfunc}) and (\ref{eq:apsym}), we
conclude that to this level of accuracy the ground state wave function at
three coinciding points is given by derivatives of the wave function of free
fermions $\Phi^{(\infty)}_0(x_1,x_2,x_3,x_4,\ldots)$ at $x_1=x_2=x_3=0$:
\begin{eqnarray}
\label{eq:gslead}
\Phi^{\!(\gamma)}_0(0,0,0,x_4,\ldots)\simeq
-\frac{1}
{(\gamma n)^3} 
\Big[\prod_{j<l}(\partial_{x_j}-\partial_{x_l})\Big]\Phi^{(\infty)}_0 .
\end{eqnarray}
Substituting Eq.~(\ref{eq:gslead}) into Eq.~(\ref{eq:g3first}) we
express the local correlator $g_3$ through derivatives of the three-body
correlation  function of free fermions. Using  Wick's theorem, the
latter  is given by  a sum of products of one-particle fermionic Green's
functions $G(x-y)= \int_{-k_F}^{k_F} dk\, e^{ik(x-y)}/2\pi$, where $k_F=\pi n$
is the Fermi wavevector. The calculation from
Eq.~(\ref{eq:g3first}) is then straightforward and we obtain ($\gamma\gg 1$)
\begin{equation}
\label{eq:g3final}
\frac{g_3(\gamma)}{n^3}=\frac{36}{\gamma^6 n^9}\!\left[\left(G''
\right)^3-G^{(4)}G''G\right]=\frac{16\pi^6}{15\gamma^6},
\end{equation}
where $G$ and its derivatives are taken at $x-y=0$.

In fact, in our problem the correlation functions $g_2$
and $g_3$ are slightly nonlocal. They are related to interparticle distances
$r\sim l_0\ll l_c$, since at smaller $r$ the relative motion of particles is
three-dimensional and the local correlators do not change. 
For large $\gamma$, at distances $\sim
l_0$ an extra (coordinate-dependent) contribution to $g_2$ is of the order of
$(nl_0)^2$ and to $g_3$ of the order of $(nl_0)^6$ (see, e.g.
\cite{MehtaRandMatr} and references therein). Under the condition $l_0\gg a$,
these contributions can be neglected as they are much smaller than the results
of Eqs.~(\ref{eq:g2large}) and (\ref{eq:g3final}), respectively.

Our method is readily generalized for the case of finite temperature
by considering the temperature-dependent Green's functions of free fermions.
For $T\ll\mu$ we obtain a small correction $\sim (T/\mu)^2$ to the
zero-temperature result. The same conclusion holds for the GP regime
($\gamma\ll 1$).

Thus, from Eq.~(\ref{eq:g3final}) we conclude that the three-body decay of 1D
trapped Bose gases is strongly suppressed in the TG regime.  Moreover,
Eq.~(\ref{g3small}) shows that even in the GP regime with $\gamma\approx
10^{-2}$, one has a 20\% reduction of the three-body rate. Thus, one also
expects a significant reduction of the three-body decay in the intermediate
regime.

For $l_0\gg a$, the recombination process in 1D trapped gases occurs at
interparticle distances much smaller than $l_0$. Therefore, the equation for
the recombination rate is the same as in 3D cylindrical Bose-Einstein
condensates with the Gaussian radial density profile. There is only an extra
reduction by a factor of $g_3/n^3$. A characteristic decay time $\tau$ is then
given by the relation $\tau^{-1}= \alpha_{3D}n_{3D}^2(g_3/3n^3)$, where
$\alpha_{3D}$ is the recombination rate constant for a 3D condensate, and
$n_{3D}=n/(\pi l_0^2)$ is the maximum 3D density. Even for $n_{3D}\sim
10^{15}$ cm$^{-3}$, the lifetime $\tau$ can greatly exceed seconds when
approaching the TG regime.  For example, this is the case for $^{87}$Rb
($\alpha_{3D}\sim 10^{-29}$ cm$^6$/s) optically trapped with $\omega_0\approx
100$ kHz ($l_0\approx 200$ \AA), assuming $L\approx 100$ $\mu$m and $N=200$.
Then one has $\gamma\approx 10$ and Eq.~(\ref{eq:g3final}) predicts a
reduction of the three-body rate by more than three orders of magnitude.

We now turn to phase coherence of a 1D Bose gas in a harmonic potential
$V(x)=m\omega^2x^2/2$.  We consider the case of $T=0$ and rely on the
hydrodynamic approach \cite{Haldane1981} in which long-wave properties of the
1D fluid are described in terms of two conjugate fields, density fluctuations
$\delta n$ and phase $\phi$.  They satisfy the commutation relation
$\left[\delta n(x),\exp\left(i\phi(x')\right)\right]=
\delta(x-x')\exp\left(i\phi(x)\right)$.  We assume the Thomas-Fermi regime and
use the local density approximation \cite{Dunjko2001,Stringari2002}: the mean
density $n(x)$ is obtained from the local equation of state
$\mu(n(x))=\mu_0-V(x)$, where $\mu(n)$ is the chemical potential for the
Lieb-Liniger problem. The density is non-zero only within the Thomas-Fermi
radius $R_{TF} =\sqrt{2\mu_0/m\omega^2}$, and the normalization condition
$\int_{-R_{TF}}^{R_{TF}} n(x) dx = N$ gives a relation between $\mu_0$ and
$N$.  Equations of motion for the fields $\delta n$ and $\phi$ follow from the
quantum Hamiltonian:
\begin{eqnarray}       
H_q=\frac{\hbar}{2\pi}\int dx \left(v_N(\pi\delta
n)^2+v_{J}(\partial_x\phi)^2\right)
=\hbar\sum_j\Omega_j{b^\dag}_{j}b_j,  \nonumber 
\end{eqnarray}
where $\Omega_j$ and $b_j$ are
frequencies and annihilation operators of elementary excitations.
The quantities $v_N(x)=(\pi\hbar)^{-1}\partial \mu/\partial n$ and
$v_J(x)=\pi\hbar n(x)/m$ determine the local sound velocity 
$\sqrt{v_N (x) v_J (x)}$ and the local Luttinger parameter
$K(x)=\sqrt{v_J (x)/v_N (x)}$. The Hamiltonian $H_q$ is a
generalization of the effective harmonic Hamiltonian of
Ref.~\cite{Haldane1981} to a non-uniform system.  

Using the density-phase representation for the field operators, we calculate
the one-particle density matrix
$g_1(x,x')=\langle\Psi^\dagger(x)\Psi(x')\rangle$ for $|x-x'|\gg l_c$. As the
density fluctuations are small, this matrix reduces to
\begin{eqnarray}     
g_1(x,x')=\sqrt{n(x)n(x')}\exp{\{-\langle
(\phi(x)-\phi(x'))^2\rangle/2\}}.    \nonumber 
\end{eqnarray}
The phase operator is given by its expansion in eigenmodes labeled by an
integer quantum number $j>0$:
\begin{equation}
\label{eq:eigen2}
\phi(x)=-i\sum_j\left(\frac{\pi v_N(0)}{2\Omega_j
R_{TF}}\right)^{\!1/2}f_j(y)b_j+\mbox{H.c.},
\end{equation}
where we have introduced a dimensionless coordinate $y=x/R_{TF}$.  The
eigenfunctions $f_j$ are normalized by the condition $\int_{-1}^{1}dy
\left(v_N(0)/v_N(y)\right)f^*_j(y)f_{j'}(y) = \delta_{jj'}$. From the
Hamiltonian $H_q$ we obtain the continuity and Euler equations which lead to
the eigenmode equation,
\begin{equation}
  \label{eq:evg}
  (1-y^2)f''_j-(2y/\beta(y))
  f'_j+(2\Omega^2/\beta(y)\omega^2) f_j = 0.
\end{equation}
The quantity $\beta(y)=d\ln\mu/d\ln n$ is determined by the local parameter
$\gamma(x)=mg/\hbar^2n(x)$.  In the TG regime, we have $\beta=2$, and in the
GP regime $\beta=1$.  The coordinate dependence of $\beta$ is smooth, and we
simplify Eq.~(\ref{eq:evg}) by setting $\beta(y)=\beta_0$, where $\beta_0$ is
the value of $\beta$ at maximum density.  This simplification has been used
\cite{Stringari2002,Combescot2002} to study the excitation spectrum of trapped
1D Bose gases. Then Eq.~(\ref{eq:evg}) yields the spectrum
$\Omega_j^2=\omega^2 (j\beta_0/2)(j+2/\beta_0-1)$, and the eigenfunctions 
$f_j(y)$ are Jacobi polynomials $P^{(\alpha,\alpha)}_j(y)$ with
$\alpha=1/\beta_0-1$.

Using Eq.~(\ref{eq:eigen2}), the mean square fluctuations of the phase
$\langle (\phi(x)-\phi(x'))^2\rangle$ are reduced to the sum over
$j$-dependent terms containing eigenfunctions $f_j$ and eigenfrequencies
$\Omega_j$.  For the vacuum phase fluctuations this sum is logarithmically
divergent at large $j$, which is similar to the high-momentum divergence in
the uniform case.  Accordingly, we introduce a cut-off $j_{max}$ following
from the condition $\hbar\Omega_j \approx \min\{\mu(x),\mu(x')\}$ and ensuring
a phonon-like character of excitations at distances $x$ and $x'$.  The vacuum
phase fluctuations have been calculated by using two approaches: numerical
summation over the eigenmodes with $f_j,\Omega_j$ from the simplified
Eq.~(\ref{eq:evg}), and quasiclassical approach assuming that the main
contribution comes from excitations with $j\gg 1$.  In the latter case, for
$x'=-x$ we obtain $\langle\left(\phi(x)-\phi(-x)\right)^2\rangle\approx K^{-1}
(x) \ln \left\{ |2x|/l_c (x)\right\}$, which is close to Haldane's result for
a uniform system \cite{Haldane1981} with the Luttinger parameter $K(x)$ and
correlation length $l_c(x)$.

\begin{figure}
  \begin{center}
    \psfrag{y}{\large$y=x/R_{TF}$} \psfrag{g}{\large$g_1(x,-x)/n(x)$}
    \psfrag{g01}{\large$\gamma_0=0.1$} \psfrag{g1}{\large$\gamma_0=1$}
    \psfrag{g10}{\large$\gamma_0=10$} 
    \psfrag{0}{\small 0}\psfrag{0.2}{\small 0.2}\psfrag{0.4}{\small 0.4}
    \psfrag{0.6}{\small 0.6}\psfrag{0.8}{\small 0.8}\psfrag{1}{\small 1} 
    \includegraphics[width=3.375in]{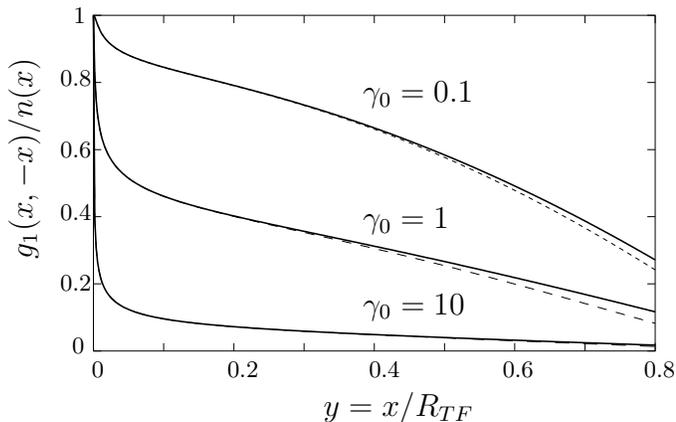}
\vspace{-0.7cm}
    \caption{The density matrix $g_1(x,-x)$ for $N=10^4$ and
      various values of $\gamma_0$. The solid curves show numerical
results, and the dashed curves the results of the quasiclassical
approach.} 
\vspace{-0.9cm}
    \label{fig:phasecorr.1}
  \end{center}
\end{figure}

The dependence of $g_1$ on the dimensionless
coordinate $y$ is governed by two parameters: $\gamma_0\equiv \gamma(0)$ and
the number of particles $N$. In Fig.~\ref{fig:phasecorr.1}, we present the
quantity $g_1(y,-y)/n(y)$ for $N=10^4$ and various values of $\gamma_0$. 
As expected, the phase coherence is completely lost in the TG regime
($\gamma_0\gg 1$). Moreover, on a distance scale $x\sim R_{TF}$ the 
coherence is already lost for $\gamma_0\approx 1$.  
Thermal fluctuations of the phase are readily
included in our scheme and will be discussed elsewhere.

In conclusion, we have found an enhanced stability of a trapped 1D Bose gas 
in the Tonks-Girardeau and intermediate regimes and described the reduction
of phase coherence in these regimes.  

We acknowledge fruitful discussions with E. Br\'ezin, R. Combescot, and V.
Kazakov and express our gratitude to Chiara Menotti for providing 
us with numerical data. This work was financially supported by the 
French Minist\`ere des Affaires Etrang\`eres, by 
the Dutch Foundations NWO and FOM, by INTAS, and by the Russian Foundation 
for Basic Research. 
Le LKB est UMR 8552 of CNRS, of ENS and of Universit\'e
P. et M.~Curie.

\end{document}